\documentclass[reprint,aps,ams,amsmath,prx,twocolumn,longbibliography]{revtex4-1}
\usepackage{graphics}
\usepackage{graphicx}
\usepackage{epsfig}
\usepackage{amsmath}
\usepackage{amssymb}
\usepackage{amsfonts}
\usepackage{bm}
\usepackage{color}
\usepackage{bbm}
\usepackage{hyperref}
\usepackage[normalem]{ulem}
\usepackage{comment}
\usepackage{soul}

\newcommand{\R}{\mathrm{Re}}
\newcommand{\Tr}{\mathrm{Tr}}

\begin{document}
\title{Admittance and critical current of nonreciprocal Josephson junctions}

\author{Tony Liu}
\affiliation{Department of Physics, University of Wisconsin-Madison, Madison, Wisconsin 53706, USA}

\author{Alex Levchenko}
\affiliation{Department of Physics, University of Wisconsin-Madison, Madison, Wisconsin 53706, USA}

\date{December 1, 2025}

\begin{abstract}
We investigate the nonequilibrium current response in diffusive superconductor-normal-metal-superconductor junctions subjected to a low-frequency AC voltage. Using a kinetic description based on the adiabatic motion of Andreev bound states, we derive a general expression for the admittance of a junction under a DC phase bias, formulated entirely in terms of the phase-dependent density of states induced by the proximity effect.
A numerical solution of the full nonlinear Usadel equations that describe the dynamics of the junction is presented. The obtained results for the admittance and the Josephson current-phase relation apply to two-dimensional planar junctions with Rashba spin-orbit coupling and an in-plane Zeeman field, as well as to Josephson junctions formed with topological insulator surface states as the normal layer. The frequency dependence of the admittance captures the crossover between the hydrodynamic and collisionless regimes, distinguished by the relation between the drive frequency and the inelastic relaxation rate in the normal region.
\end{abstract}

\maketitle

\section{Introduction}

Josephson junctions driven out of equilibrium by external voltages exhibit complex dynamical responses. In the simplest case of a tunnel-type junction, such as a superconductor-insulator-superconductor (SIS) junction, the current response can be calculated for an arbitrarily time-dependent applied voltage \cite{LO:JETP67}. In the limiting case, the governing equations for the superconducting phase \cite{AES:PRB84}, when coupled to the dissipative electromagnetic environment, can be reduced to a simple circuit description -- the resistively and capacitively shunted junction (RCSJ) model, which is widely used in modeling superconducting electronic devices \cite{Likharev:Book91}.

When the insulating barrier of a junction is replaced by a normal metal, forming a superconductor–normal metal–superconductor (SNS) junction, the dynamical properties of the device are governed by the interplay between the proximity effect (PE) \cite{Likharev:RMP79,GKI:RMP04} and multiple Andreev reflections (MAR) \cite{Zaitsev:PRL98,Bazuglyi:PRB00,Cuevas:PRB06}. The proximity effect induces a spectral gap in the normal region. For a long disordered normal layer, the size of this gap is of the order of the Thouless energy \cite{Belzig:PRB96}, which is determined by the dwell time required for electrons to diffuse across the layer between the superconducting reservoirs and establish phase-coherent transport. On top of that, multiple Andreev reflections at the superconducting interfaces along with the inelastic processes generate a highly nontrivial energy distribution of electrons in the normal layer. As a result, calculating even a linear-response coefficient, such as the admittance of the junction, poses major technical challenges due to the nonperturbative nature of both PE and MAR. 

The AC response of diffusive SNS junctions has been investigated theoretically in numerous works \cite{Lempitskii:JETP83,Volkov:PRL96,Zhou:JETPLett97,Argaman:SM99,Brinkman:PRB03,Virtanen:PRL10,Virtanen:PRB11,Ferrier:PRB13,Tikhonov:PRB15} and measured experimentally including devices where the normal layer was a topological material \cite{Lehnert:PRL99,Chiodi:SR2011,Dassonneville:PRL13,Stehno:PRB25}. In particular, the frequency-dependent admittance
~\footnote{We recall that, in addition to the admittance of the SNS junction $\Upsilon(\omega)$ one often considers its linear susceptibility $\chi(\omega)$. These two quantities are related in the geometry of a ring containing an embedded SNS junction, where the phase difference is controlled by the magnetic flux $\Phi$. Since the electromotive force driving the current involves the time derivative of $\Phi$, the susceptibility, defined as $\chi(\omega)=\partial_\Phi J$, is related to the admittance by an extra power of frequency, $\chi(\omega)=i\omega\Upsilon(\omega)$.}
\begin{equation}
\Upsilon(\omega, \phi_0 ) = \frac{ J(\omega, \phi_0)}{ U(\omega)}, 
\end{equation}
which characterizes the linear current response $J$ of a junction with phase bias $\phi_0$ to an applied voltage $U$, has been calculated across a variety of temperature and frequency regimes. Crucially, the superconducting proximity effect causes 
$\Upsilon$ to deviate from the admittance of a normal metallic bridge (wire) even in the DC limit at finite temperature owing to the energy dependence of the diffusivity of quasiparticles \cite{Artemenko:SSC79,Nazarov:PRL96,Stoof:PRB96}.

For a long junction, $E_T\ll\Delta(T)$, where $\Delta(T)$ is the superconducting energy gap at temperature $T$, and $E_T=D/L^2$ is the Thouless energy of a disordered junction of length $L$ and diffusion constant $D$, the admittance was first described theoretically by Lempitskii \cite{Lempitskii:JETP83}, who derived the following analytical expression:
\begin{align}\label{Lempitskii}
\Upsilon(\omega, \phi_0) = \Upsilon_N \left( \frac{E_T}{ T} \right) \left( \frac{\tau_{in} E_T }{1 - i\omega \tau_{in} }  \right)Q(\phi_0).
\end{align}
In this formula, $\tau_{in}$ denotes the inelastic scattering time of electrons in the normal layer, $\Upsilon_N$ is the normal-state admittance of the junction, and $Q(\phi_0)$ is a dimensionless function that can be expressed in terms of Green's functions of a disordered superconductor. The applicability of Eq.~\eqref{Lempitskii} requires the adiabatic regime, $\omega<E_T$; nevertheless, it captures the system’s response across the crossover from the hydrodynamic limit, $\omega\tau_{in}\ll1$, to the collisionless limit, $\omega\tau_{in}\gg1$. Originally, Eq. \eqref{Lempitskii} was derived under the additional assumption of high temperature, $T>E_T$. However, improved computations (see, e.g., Refs. \cite{Virtanen:PRB11,Tikhonov:PRB15}) have relaxed this restriction. It should also be noted that MAR processes are not essential in the regime of small applied voltages, which is another limiting assumption underlying Eq. \eqref{Lempitskii}. 

At higher frequencies, $\omega>E_T$, the admittance of the junction reduces to that of two SN junctions connected in series. In this incoherent limit, $\Upsilon$ becomes independent of $\phi_0$. By contrast, in the adiabatic limit the admittance exhibits a strong $\phi_0$ dependence, which persists even at high temperatures, $T>E_T$, where phase coherence would normally be expected to weaken. The crucial point and the practical value of Eq. \eqref{Lempitskii} and its generalizations is that they provide a useful tool for inferring the dynamical properties of Andreev levels in the junction through their impact on the conductive properties of the proximity-modified normal region.

Motivated by the recent interest in nonreciprocal Josephson junctions and superconducting diode effects \cite{Jiang:NP2022,Nadeem:NRP2023,Ma:APR25,Shaffer:2025}, we extend Lempitskii’s result to the case of noncentrosymmetric SNS junctions and investigate their dynamical AC response. In contrast to earlier approaches based on Green’s function techniques, we describe the low-frequency dynamics of the junction in terms of the quasiparticle distribution function $n(\epsilon,t)$, subject to the adiabatic motion of instantaneous energy levels induced by the phase dynamics. Using the framework developed in Refs. \cite{Smith:PRB21,Liu:PRB24}, which captures the dynamics of SNS junctions in the low-frequency regime, we derive a general expression for $\Upsilon(\omega,\phi_0)$ in terms of the phase-dependent density of states in the junction. Finally, we apply these results to a diffusive planar SNS junction with spin-orbit coupling and an in-plane Zeeman field, and describe the phase dependence of the  admittance for different parameters of the model. Our results are also applicable to superconductor-topological insulator-superconductor (S-TI-S) junctions in the diffusive limit. In parallel, we investigate the anomalous Josephson current–phase relation in these devices as part of our broader analysis.

\section{Kinetic description of adiabatic dynamics in SNS junctions}

Consider an SNS junction subjected to an external voltage, $U(t) = \int d\omega U(\omega) e^{i\omega t}$. 
The phase difference $\phi(t)$ across the junction then evolves in time according to the Josephson relation
\begin{equation}\label{eq:josephson}
\dot{\phi}\equiv \frac{d\phi}{dt}  = 2eU(t).
\end{equation}
In the presence of an AC voltage, the phase difference oscillates around a DC phase bias $\phi_0$. Because the Andreev bound states in the junction are sensitive to $\phi(t)$, 
the phase dynamics induced by the external drive influence the quasiparticle kinetics and complicate the current response.

Due to Andreev reflection at the normal metal-superconductor interfaces, sub-gap quasiparticles with energies $\epsilon < \Delta$ are confined within the normal region of the junction, and their spectrum depends sensitively on the superconducting phase difference $\phi$. When an external voltage is applied, in accordance with Eq. \eqref{eq:josephson} the resulting phase dynamics drive temporal variations of these energy levels, inducing a spectral flow. In the limit of small voltages, the quasiparticles occupying these levels follow the levels adiabatically, leading to the formation of a nonequilibrium quasiparticle distribution.

The kinetic equation for the distribution $n(\epsilon, t)$ can be phenomenologically derived by considering two continuity equations.
The first equation describes the conservation of the number of energy levels,
\begin{align}\label{cont1}
\partial_t \nu (\epsilon, \phi(t)) + \partial_\epsilon \left[ v_\nu \nu\left(\epsilon, \phi(t) \right) \right] = 0 .
\end{align}
Here $\nu\left(\epsilon, \phi(t) \right) $ is the phase dependent density of states in the junction and $v_\nu\left(\epsilon, \phi(t) \right) $ is a phenomenologically introduced parameter representing the ``velocity" of the levels in energy space.
The second equation represents the conservation of quasiparticles,
\begin{align}\label{cont2}
\partial_t \left[ n(\epsilon, t) \nu (\epsilon, \phi(t)) \right]  + \partial_\epsilon \left[ n(\epsilon, t) v_\nu\left(\epsilon, \phi(t) \right) \nu\left(\epsilon, \phi(t) \right) \right]  \nonumber \\
= I \left[n(\epsilon, t) \right] .
\end{align}
Here $I\left[ n(\epsilon, t) \right]$ is the collision integral. Although its exact form is generally complex and depends on the microscopic scattering mechanisms 
(e.g., electron-phonon interactions), for simplicity we treat it within the relaxation-time approximation, 
\begin{equation}
I\left[ n(\epsilon, t) \right] =-\frac{\delta n(\epsilon, t) }{\tau_{in}},
\end{equation}
where $\delta n(\epsilon, t) = n(\epsilon, t) - n_F(\epsilon)$ is the deviation of the quasiparticle distribution $n(\epsilon, t)$ from the equilibrium Fermi function $n_F(\epsilon)$ that relaxes over the typical time 
$\tau_{in}$~\footnote{In a superconductor, the electron–phonon relaxation rate at not too low temperatures is of the same order as in a normal metal, $\tau_{in}\propto \Theta^2_D/T^3$,
where $\Theta_D$ is the Debye energy. This behavior arises because phonon emission does not involve a significant change in the direction of the electron momentum.}.  
By substituting Eq.~\eqref{cont1} in \eqref{cont2} and using the Josephson relation, we get the kinetic equation for $n(\epsilon, t)$,
\begin{equation}\label{EQ:N_DOT}
 \partial_{t} n (\epsilon, t )+2e U(t) V_{\nu} \left(\epsilon, \phi(t) \right)\,  \partial_\epsilon  n(\epsilon, t ) = -\frac{\delta n(\epsilon, t) }{\tau_{in}}.
\end{equation}
Here we have defined 
\begin{equation}\label{eq:level_sensitivity}
V_{\nu} \big(\epsilon,\phi \big) \equiv -\frac{1}{\nu (\epsilon, \phi)}  \int_{0}^{\epsilon} d{\varepsilon}  \frac{\partial \nu (\varepsilon, \phi )}{\partial \phi},
\end{equation}
which can be interpreted as the sensitivity of the energy levels to a change in the phase difference. 
The second term on the left-hand-side of Eq.~\eqref{EQ:N_DOT} is a source term, which describes the influence of the spectral flow of energy levels on the distribution function.

In addition to the kinetic equation, we will also need an expression for the current.
In the adiabatic regime, contributions to the current associated with transitions between energy levels can be neglected and the current across the junction can be expressed as follows
\begin{equation}\label{EQ:CURRENTEUL}
J(t) = -2e \int_{0}^{\infty} d \epsilon \nu \big(\epsilon, \phi(t) \big)V_{\nu}\big( \epsilon, \phi(t) \big) \delta n(\epsilon,t)+  J_{s} \big( \phi(t) \big).
\end{equation}
Here $\nu(\epsilon, \phi)$ is the total density of states and  $J_s(\phi)$ is the equilibrium supercurrent.
The supercurrent contains contributions from both the ground state and the equilibrium quasiparticle excitations. It can be written as  
 \begin{equation}\label{supercurrent}
J_s(\phi) = -2e \int^\infty_0 d\epsilon \tanh\left(\frac{\epsilon}{2T}\right) \nu(\epsilon,\phi) V_\nu(\epsilon,\phi), 
\end{equation}
Although Eqs.~\eqref{EQ:N_DOT}--\eqref{EQ:CURRENTEUL} can be derived rigorously using a method involving Green's functions, 
we will not discuss the details of this derivation in this article (see e.g. Refs. \cite{Liu:PRB24,Pesin:PRL08} for further technical details). 

\section{General formula for the admittance of diffusive SNS junctions}

As is evident from the kinetic theory formulated in the previous section, the total current in a junction [Eq. \eqref{EQ:CURRENTEUL}], including both dissipative and superflow contributions, is highly nonlinear in the applied time-dependent voltage, even within the simplified adiabatic approximation. Consequently, obtaining the full current-voltage characteristic analytically remains a challenging task. We therefore focus on the linear-response regime, where a closed-form analytical expression for the admittance can be derived. The exact functional dependence of the admittance on the phase bias and other control parameters of the model must be evaluated numerically. We demonstrate this by solving the Usadel equation without assuming a weak proximity effect in the following section. In particular, we account for all nonperturbative features, such as the minigap. The reduction of the problem to a spectral calculation makes our approach more controlled than a fully numerical solution of the time-dependent Usadel equation.

When a junction with phase bias $\phi_0$ applied across superconducting reservoirs is subject to an AC voltage, the phase dynamics are given by the solution of the Josephson equation \eqref{eq:josephson} in the form 
\begin{align}\label{dynamicchi}
\phi(t) = \phi_0 + \int d\omega  \left( \frac{2e U(\omega)}{ i\omega} \right) e^{i\omega t}.
\end{align}
Substituting this into Eq.~\eqref{EQ:N_DOT} and linearizing the resulting kinetic equation with respect to $U(\omega)$, which amounts to replacing $\phi(t)\to\phi_0$ in all the functions describing the spectral flow, we obtain
\begin{align}
\left( \partial_{t} + \frac{1}{\tau_{in}}\right)  \delta n (\epsilon, t ) = -2eU(t)V_{\nu} (\epsilon, \phi_0 )\,  \partial_\epsilon  n_F(\epsilon) .
\end{align}
Solving this equation by the Fourier transform we have 
\begin{align}
\delta n(\epsilon, t) =-\int d\omega \left( \frac{2eU(\omega) \tau_{in} }{1 + i\omega \tau_{in} } \right) \partial_\epsilon n_F(\epsilon) V_\nu(\epsilon, \phi_0)  e^{i\omega t}.
\end{align}
Next, substituting the result into the expression for the current Eq.~\eqref{EQ:CURRENTEUL}, we get
\begin{align}\label{eq:Jt}
J(t) &= J_s(\phi_0)  + \int d\omega \partial_\phi J_s(\phi_0) \left( \frac{2eU}{i\omega} \right) e^{i\omega t} \nonumber \\
& +\int d\omega \left( \frac{4e^2U(\omega) \tau_{in} }{1 + i\omega \tau_{in} } \right)e^{i\omega t} \nonumber \\
& \,\,\,\,\, \times  \int^\infty_0 d \epsilon \nu(\epsilon, \phi_0) V_\nu^2 (\epsilon, \phi_0) \partial_\epsilon n_F(\epsilon) .
\end{align}
\normalsize
The first two terms describe the variation of the equilibrium supercurrent as $\phi$ oscillates, and they do not contribute to the dissipation in the junction.
The real part of the admittance is then given by the following formula,
\begin{align}\label{admittance}
\R \Upsilon (\omega, \phi_0) = \Upsilon_N \left( \frac{E_T}{T} \right)\left( \frac{ \tau_{in} E_T  }{1 + \left( \omega \tau_{in} \right)^2 } \right) Q(\phi_0,T).
\end{align}
Here $\Upsilon_N = e^2 \nu_N E_T$ is the admittance of the normal metal region, and we have defined the dimensionless function
\begin{align}\label{Q}
Q(\phi_0, T)=\int^\infty_0 \frac{\nu(q E_T, \phi_0)d q}{\nu_N\cosh^2\frac{qE_T}{2T}} &
 \left(\frac{V_\nu (q E_T, \phi_0)}{E_T} \right)^2 .
\end{align}
For the case of long diffusive junctions, it can be shown that $Q(\phi_0)$ reduces to the known result and thus Eq.~\eqref{admittance} is consistent with the Lempitskii formula \eqref{Lempitskii}.
The main difference between this result and the Lempitskii formula is that the dimensionless function $Q(\phi_0)$ is expressed in terms of the phase dependent density of states $\nu(\epsilon, \phi_0)$ rather than superconducting 
Green's functions.

Another contribution to the admittance arises from the AC supercurrent. Indeed, it follows from Eq. \eqref{eq:Jt} that the corresponding part takes the form 
\begin{equation}\label{eq:Ys}
\Upsilon_s(\omega,\phi_0)=\frac{2e}{i\omega}\partial_\phi J_s(\phi_0).
\end{equation}
From Eq.~\eqref{supercurrent} we can obtain the Josephson current-phase relation (CPR). In the scattering theory of mesoscopic systems, CPR is usually expressed in terms of separate contributions of sub-gap Andreev bound states and the continuum above the gap \cite{Beenakker:1992}. In contrast, Eq. \eqref{supercurrent} gives an alternative representation. Using Eq. \eqref{eq:level_sensitivity} we express CPR in terms of the density of states 
\begin{align}\label{eq:J-CPR}
&J_s(\phi_0)=\frac{2\Upsilon_NE_T}{e}K(\phi_0) \nonumber,  \\ 
&K(\phi_0)=\int\limits^{\infty}_{0}dq\int\limits^{q}_{0}dq'\tanh\left(\frac{qE_T}{2T}\right)\frac{\partial\nu(q'E_T,\phi_0)}{\nu_N\partial\phi_0}.
\end{align} 

Since we have not assumed any specific form of the density of states, Eqs.~\eqref{admittance}–\eqref{eq:J-CPR} remain fully general, in direct analogy to the corresponding formula for the linear conductivity \cite{Liu:PRB24}. As long as the applied voltage and frequency are sufficiently small, this expression for the admittance holds irrespective of the details of the normal region, including the strength of disorder, the form of spin–orbit coupling, or the presence of an external magnetic field. Consequently, the problem of determining the linear admittance in nonreciprocal junctions reduces to calculating the density of states. This conclusion represents a key result of this work.

In noncentrosymmetric junctions, we introduce a parameter $\Xi(\phi_0)$ that quantifies the degree of asymmetry in the dissipative response with respect to the phase bias,
\begin{align}\label{deltaupsilon}
\Xi(\phi_0) & = \R \left( \frac{\Upsilon (\omega, \phi_0)   - \Upsilon (\omega, -\phi_0)  }{\Upsilon (\omega, \phi_0)   + \Upsilon (\omega, -\phi_0)  } \right) \nonumber \\
 &=  \frac{Q(\phi_0) - Q(-\phi_0) }{Q(\phi_0) + Q(-\phi_0) }.
\end{align}
This definition closely parallels that of the diode efficiency recently introduced in the context of the superconducting diode effect. Whereas diode efficiency measures the asymmetry of the critical supercurrent 
$J_s$, the parameter $\Xi$ captures the asymmetry of the dissipative current in the junction. 
Importantly, in junctions possessing either time-reversal or inversion symmetry, the density of states satisfies 
$\nu(\epsilon, \phi) = \nu(\epsilon, -\phi)$, consequently, the asymmetry parameter vanishes, $\Xi = 0$.

\begin{figure}[t!]
\includegraphics[width=0.4\textwidth]{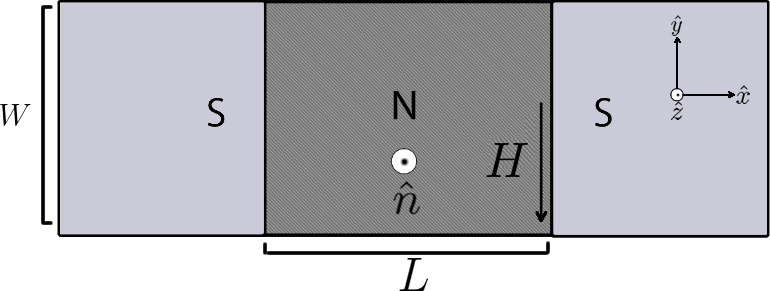}
\caption{Top down view of a planar SNS junction. The junction is aligned along the $ \hat{\bf x}$ direction, there is a parallel magnetic field ${\bf H}$ directed in the $\hat{\bf  y}$ direction, 
and there is an out of plane vector $\hat{\bf n}$ pointing in the $\hat{\bf z}$ direction from the spin-orbit effect that breaks inversion symmetry.}
\label{SNSplanar}
\end{figure}
\section{Application to a junction with spin-orbit coupling in magnetic field}

As an illustrative example of a junction that breaks the necessary symmetries to allow for a nonreciprocal response,
we consider a diffusive planar SNS junction with an in-plane magnetic field ${\bf H}$ and Rashba spin-orbit coupling (SOC) in the normal region. The normal region is described by the Hamiltonian
\begin{align} \label{hamiltonian} 
	H &= {\bf p}^2/2m - E_{F} + \beta^{\alpha i} p^i \sigma^\alpha 
	 +  V_{imp}({\bf r}) + g \mu_B {\bf H} \cdot {\boldsymbol \sigma} .
\end{align}
Here $E_{F}$ is the Fermi energy, $m$ is the electron effective mass, $\sigma_i$ are the Pauli matrices in spin space, $g$ is the gyromagnetic factor, $\mu_B$ is the Bohr magneton, and $V_{imp}({\bf r})$ is the random impurity potential.
For Rashba spin-orbit coupling $\beta^{\alpha i} = \alpha_R \epsilon^{\alpha i j}\hat n^j$, where $\hat n$ is a unit polar vector and parameter $\alpha_R$ has units of velocity.
Since the magnetic field is chosen to be parallel to the plane, see Fig. \ref{SNSplanar}, it enters the Hamiltonian, Eq.~\eqref{hamiltonian},  only via the Zeeman term.
For simplicity, we consider junctions whose length greatly exceeds their width, $L \gg W$, allowing us to treat them in the quasi-one-dimensional limit. In addition, we assume that the junction size is much larger than both the mean free path and the spin-relaxation length, $L \gg \{l, l_{so}\}$, ensuring that transport is fully diffusive.

\subsection{Usadel equations}
In general, the average density of states $\nu(\epsilon, \phi)$ of disordered superconducting systems can be obtained by solving Usadel equations \cite{Usadel:PRL1970},
which in the presence of linear spin-orbit and a Zeeman fields corresponding to the Hamiltonian Eq. \eqref{hamiltonian} are given by \cite{Tokatly:PRB17}
\begin{align}
\label{eq:UsadelNambuSpin}
\left[\epsilon\hat \tau_3 - \Delta({\bf r}) \hat \tau_1 + A_0 \hat \tau_3, \check g \right] - i  \partial_{\bf r}^k  \left( D\check g \partial_{\bf r}^k  \check g \right)  \nonumber \\
- {\frac{i}{2m}} \left\{ A_{k}({\bf r}), \nabla_{\bf r}^k  \check g \right\} = 0.
\end{align}
Here $\check g(\epsilon, {\bf r} , \phi)$ is the isotropic part of the quasi-classical Green's function,  $D = \tau_{el} l/2$ is the diffusion coefficient for elastic scattering on impurity potential $V_{imp}({\bf r})$. The notations $[\cdot, \cdot]$ and $\{ \cdot, \cdot\}$ are used to denote the commutator and anti-commutator operator respectively. $\Delta({\bf r})$ is the superconducting order parameter given by 
\begin{equation}\label{Delta}
\Delta ({\bf r}) = 
\begin{cases}
\Delta e^{-i\phi/2}, & \quad x < - \frac{L}{2},\\
0, & \quad -\frac{L}{2} < x < \frac{L}{2},\\
\Delta e^{i\phi/2}, & \quad x> \frac{L}{2}.
\end{cases} 
\end{equation}
The effect of the spin-orbit and Zeeman fields are captured by the fictitious SU(2) gauge fields, which are defined as  
\begin{align}
A_0 = & \,  - g \mu_B H^i \sigma^i, \qquad A_k =  {-m \beta^{i k} \sigma^i}.
\end{align}
Finally, we have introduced the gauge covariant derivative $\partial_{\bf r}^k   = \nabla^k_{\bf r} - i[A_k, \cdot] $.

The Green's function $\check g$ is a 4$\times$4 matrix in Nambu-spin space subject to the normalization condition $\check g^2 = 1$.
The density of states can then be obtained from the Green's functions using the identity,
\begin{align}\label{DOSidentity}
\nu(\epsilon, \phi) = \frac{\nu_N}{2L} \int^{L/2}_{-L/2} d x \R\big[ \Tr \big( \check \tau_3 \check g(\epsilon, x, \phi ) \big) \big],
\end{align}
where and $\Tr$ is the trace over Nambu-$\tau$ and spin-$\sigma$ Pauli matrices respectively.

The boundary equation for Eq.~\eqref{eq:UsadelNambuSpin} depends on the transparency of the NS interfaces. 
For simplicity, we will consider the case where the NS interfaces are fully transparent, such that the $\check g$ 
evaluated at the boundaries is equal to Green's function in the bulk of the superconducting electrodes. 
For energies which are small compared to the bulk gap, $\{T,E_T\}\ll\Delta$, the effective boundary condition is given by
\begin{align}\label{boundarycondition}
\check g\left( \pm \frac{ L}{2} \right) = 
\begin{pmatrix}
0 & e^{\pm  \frac{i\phi}{2} } \\
e^{\mp \frac{i\phi}{2}  } & 0 \\
\end{pmatrix}
\sigma_0.
\end{align}
In this approximation, we assume that the superconducting gap drops abruptly to zero at the NS interfaces, neglecting small corrections that would arise from enforcing self-consistency.

\begin{figure}[t!]
\includegraphics[width=0.5\textwidth]{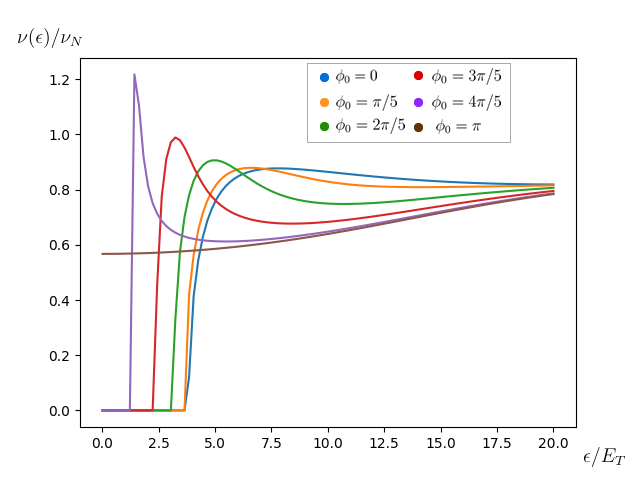}
\caption{Numerically calculated density of states in a diffusive SNS junction from the solution of Usadel equations \eqref{eq:Usadel} plotted for various phase differences $\phi_0$. 
The energy is normalized by the Thouless energy $E_T$.}
\label{Fig:DOS}
\end{figure}

\subsection{Perturbative solution in magnetic field}

The Usadel equations \eqref{eq:UsadelNambuSpin} can be greatly simplified in the case of weak spin-orbit coupling, $\alpha_Rp_F\ll\tau^{-1}_{el}$, and relatively weak magnetic field, $h<\tau^{-1}_{so}$,  
where  $h = g\mu_B H$ is the Zeeman energy and $\tau_{so} = (p_F^2 \alpha_R^2 \tau_{el})^{-1} $  is the Dyakonov-Perel spin relaxation time \cite{DyakonovPerel}. 
The details of this procedure are given in Ref.~\cite{Liu:PRB24} and we briefly summarize the main steps required for our analysis of the admittance calculation. 

In this approach, the Green’s function 
$\check g$ is decomposed into spin-singlet and spin-triplet components,
\begin{align}\label{eq:gs-gt}
\check g = \hat g_s  + \hat g_t^k \check \sigma^k,
\end{align}
with the implicit assumption that $\hat g_s\gg\hat g_t$. In this approximation the normalization conditions becomes $\hat{g}^2_s=1$ and $\{\hat{g}_s,\hat{g}^\alpha_t\}=0$. 
Inserting Eq. \eqref{eq:gs-gt} into Eq. \eqref{eq:UsadelNambuSpin} one finds two coupled equations for $\hat g_s $ and $\hat g_t$. The equation for the triplet component can be solved in the form 
\begin{equation}
\hat{g}_t=-i\tau_{so}\left(\hat{g}_s[h^\alpha\hat{\tau}_3,\hat{g}_s]-i\beta^{\alpha k}\hat{g}_s\nabla^k_{\bf r}\hat{g}_s\right),
\end{equation}
which is valid provided an additional assumption $E_T\ll \tau^{-1}_{so}$ is satisfied. This, in turn, leads to a closed equation for the singlet component of the form
\begin{align}\label{eq:Usadel-gs}
&D\nabla^k_{\bf r}(\hat{g}_s\nabla^k_{\bf r}\hat{g}_s)+i[\epsilon\hat{\tau}_3,\hat{g}_s]-\tau_{so}h^2[\hat{\tau}_3,\hat{g}_s[\hat{\tau}_3,\hat{g}_s]]\nonumber \\  
&+i\tau_{so}\beta^{\alpha k}h^\alpha([\hat{\tau}_3,\hat{g}_s\nabla^k_{\bf r}\hat{g}_s]+\nabla^k_{\bf r}(\hat{g}_s[\hat{\tau}_3,\hat{g}_s]))=0.
\end{align}

In the following, we choose a conventional parameterization for $g_s$ in terms of two complex variables $(\theta, \varphi)$,
\begin{align}
	\hat g_s (\epsilon,x) = \left(\begin{array}{cc}
		\cos\theta(\epsilon,x) & \sin\theta(\epsilon,x) e^{i \varphi(\epsilon,x)}\\
		\sin\theta(\epsilon,x) e^{-i \varphi(\epsilon,x)} & -\cos\theta(\epsilon,x)
	\end{array}\right),
\end{align}
which explicitly resolves the nonlinear constraint imposed on the Green's function by its normalization condition.
Unpacking then the Nambu components of Eq. \eqref{eq:Usadel-gs} for the quasi-1D geometry and field configuration specific to Fig. \ref{SNSplanar}
one arrives at two coupled nonlinear equations that completely define the singlet component of the Green's function   
\begin{subequations}\label{eq:Usadel}
\begin{align}
\label{eq:ParamEq1}
	\partial_x  \left[  \big(D\partial_x \varphi + 2\tau_{so} \alpha_R h \big) \sin^2\theta  \right] =0,
\end{align}
\begin{align}
\label{eq:ParamEq2}
 D\partial^2_x \theta  + 2i\epsilon\sin\theta -2\tau_{so}h^2\sin2\theta\nonumber \\ 
- \frac{D}{2}\sin 2\theta \bigg(\partial_x\varphi  + \frac{2\tau_{so} \alpha_R h  }{D} \bigg)^2
  = 0.
\end{align}
\end{subequations}
In these variables, the boundary condition Eq.~\eqref{boundarycondition} is given by
\begin{align}\label{BC}
	\theta (\epsilon, \phi, \pm L/2) &= \frac{\pi}{2}, \quad 
 \varphi(\epsilon, \phi, \pm L/2 ) =  \pm \frac{\phi_0}{2}.
\end{align}
The expression for the density of states Eq.~\eqref{DOSidentity} is equivalently given by
\begin{align}\label{DOS3}
	\nu(\epsilon,\phi ) &=  \frac{\nu_N}{L} \int^{L/2}_{-L/2}  d {x}  \R \left[\cos\theta(\epsilon, \phi_0, x )\right].
\end{align}

\begin{figure}[t!]
\includegraphics[width=0.5\textwidth]{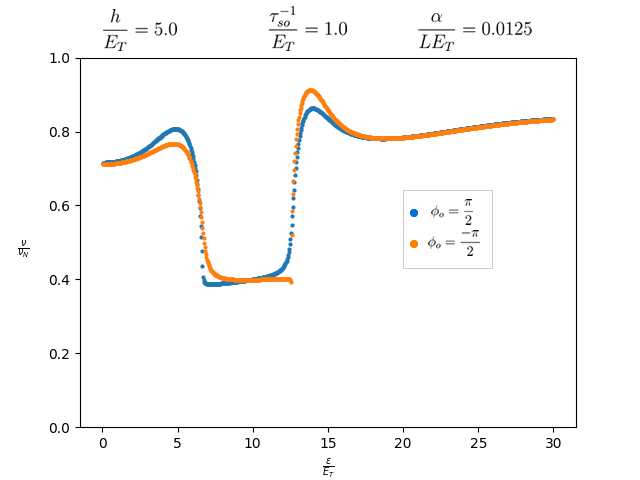}
\includegraphics[width=0.5\textwidth]{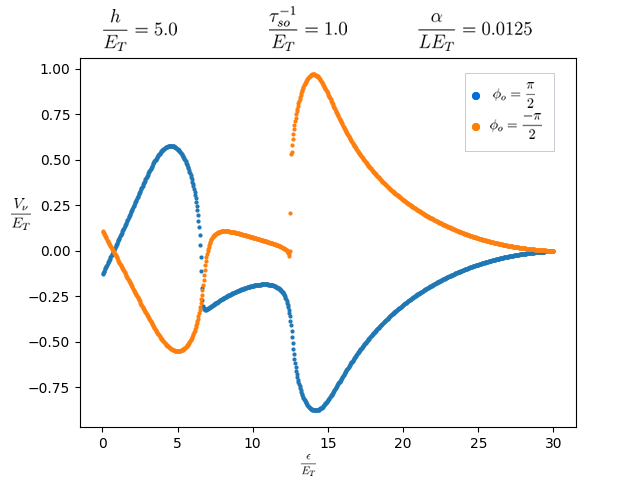}
\caption{The top panel shows numerically calculated density of states in a diffusive SNS junction plotted for a sufficiently high magnetic field $h=5E_T$ at two representative values of the phase $\phi_0$ across the junction. 
The values of other control parameters related to the strength of spin-orbit are listed on the plot. The bottom panel shows sensitivity of the energy
levels $V_\nu(\epsilon,\phi_0)$ to a change in the phase difference plotted for the same set of parameters.}
\label{Fig:DOS-h}
\end{figure}

\subsection{Linear order estimates}

We make an observation that the terms in Eqs.~\eqref{eq:ParamEq1}-\eqref{eq:ParamEq2} which are linearly proportional to $h$ can be absorbed into a shift in the local phase $\varphi(\epsilon, x)$.
Indeed, this is achieved by introducing a gauge transformation with a change of variables $(\theta, \varphi) \rightarrow (\theta, \chi)$, where
\begin{equation}
\chi (\epsilon, \phi, x) = \varphi(\epsilon, \phi,x)  + \left( \frac{2\tau_{so} \alpha_R h }{D}\right) (x+L/2),
\end{equation}
As a result, with the accuracy up to linear in $h\tau_{so}<1$ order, the Usadel equations \eqref{eq:Usadel} written in terms of $\chi$ 
are equivalent to Usadel's equations for a junction with $h =0$ and $\alpha_R = 0$, biased at phase difference of $\phi_0 + \delta\phi$, 
where the additional phase shift is given by 
\begin{equation}
\delta \phi =  \frac{2\tau_{so}\alpha_R h L }{D}.
\end{equation}
This means that the solution for $\theta$ is equal to the zero field solution with a phase shift of $\delta \phi$.
Thus, the density of states, which depends only on $\theta(\epsilon, \phi)$, can be expressed in terms of the zero field density of states $\nu_0(\epsilon,\phi)$,
\begin{align}\label{nuestimateh}
\nu(\epsilon,\phi, {\bf H}) &=\nu_0 (\epsilon,\phi + \delta \phi). 
\end{align}
Since $Q(\phi_0,{\bf H} )$ is defined as a functional of $\nu(\epsilon, \phi_0, {\bf H})$, it is shifted by the same phase,
\begin{align}\label{Qshift}
Q(\phi_0, {\bf H}) = Q(\phi_0 + \delta \phi, 0).
\end{align}
This conclusion is reminiscent of that concerning the anomalous Josephson effect in SNS junctions, where the current-phase relation preserves its functional form but acquires a phase shift induced by spin–orbit coupling and the applied field \cite{Buzdin:PRL08,Bergeret:EPL15,Hasan:PRB22}.

\begin{figure*}[t!]
\includegraphics[width=0.3\linewidth]{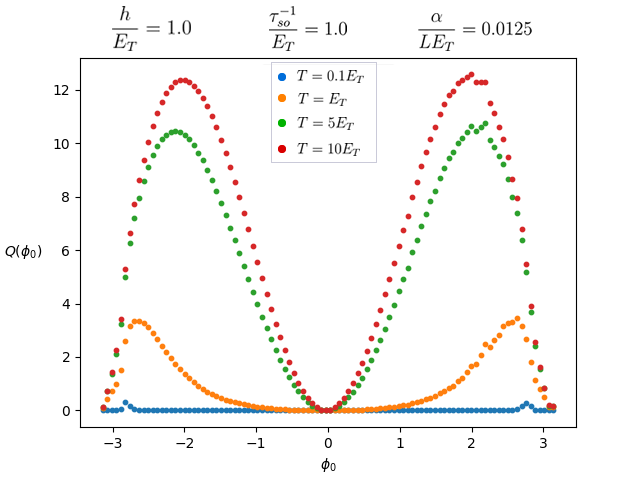}
\includegraphics[width=0.3\linewidth]{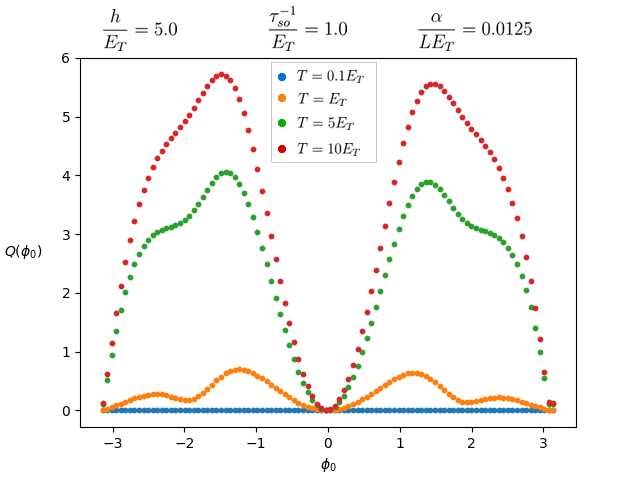}
\includegraphics[width=0.315\linewidth]{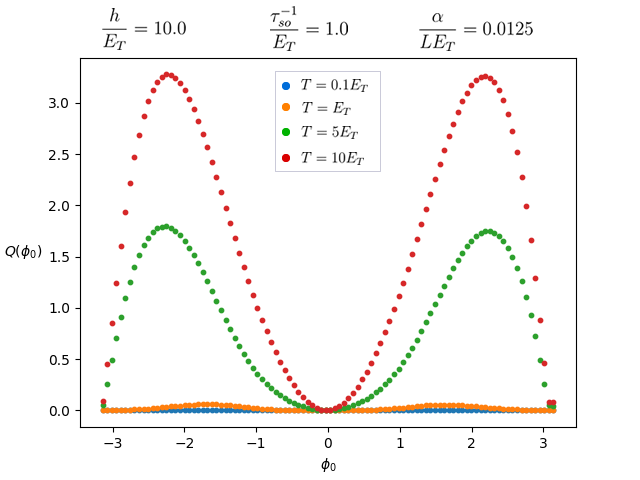}
\caption{Variation of the dimensionless admittance function of a long SNS junction as a function of temperature and magnetic field. Parameters of the plot are shown for each line. }
\label{Fig:Q}
\end{figure*}

Since the Usadel equation remains nonlinear even at $H=0$, it is not possible to obtain a fully closed analytic expression for $Q(\phi_0, 0)$. 
Nevertheless, because zero-field solutions of the Usadel equation in diffusive SNS junctions have been extensively studied (see, e.g., Refs. \cite{Belzig:PRB96,Zhou:JLTP98,Levchenko:PRB08,Fominov:AoP23}) and the main qualitative features of $\nu_0(\epsilon, \phi)$ are well established, we can estimate the magnitude of both $Q(\phi_0, 0)$ and $\Xi(\phi_0, {\bf H})$ without fully resorting to numerics.

The main feature of $\nu_0(\epsilon, \phi)$ is the presence of a phase-dependent minigap $E_g(\phi)$, which has a size of the order of the Thouless energy $E_T$ and closes at $\phi = \pi$, see Fig.~\ref{Fig:DOS}. 
At energies above the minigap, $\epsilon \gg E_T$, the density of states asymptotically approaches $\nu_N$ and becomes weakly sensitive to the phase difference $\phi$.
If the temperature is above the gap $E_T \lesssim T$, the dominant contribution to $Q(\phi,0)$ comes from the energy interval $\epsilon \lesssim E_T$, where the level sensitivity has a characteristic size $V_\nu \sim E_T$ and $\nu_0 \sim \nu_N$.
Substituting these crude estimates in Eq.~\eqref{Q}, we get 
\begin{equation}
\Xi(\phi_0) \approx \frac{\partial_\phi Q(\phi_0, 0)  }{Q(\phi_0, 0)} \delta \phi \sim \frac{\alpha_R\tau_{so}}{L}\frac{g\mu_BH}{E_T}f(\phi_0)
\end{equation}
where $f(\phi_0)\sim1$ is the dimensional function.  

\begin{figure*}[t!]
\includegraphics[width=0.3\linewidth]{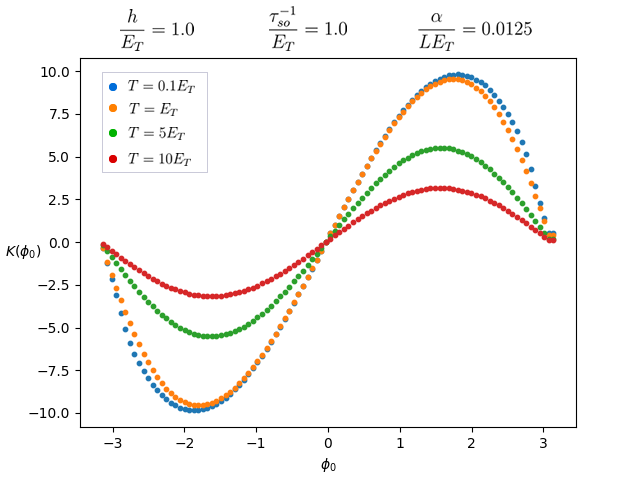}
\includegraphics[width=0.3\linewidth]{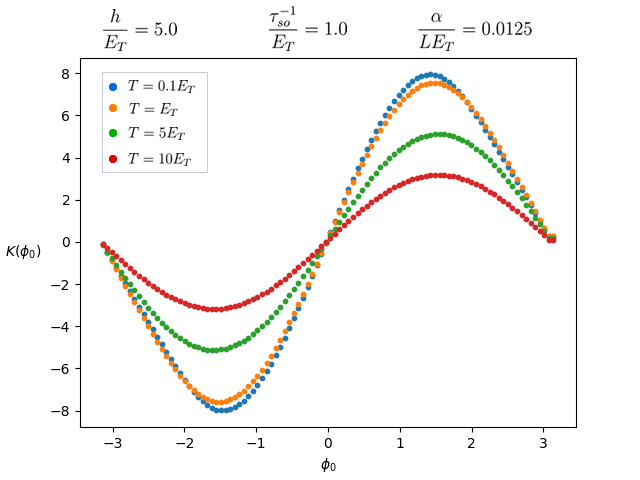}
\includegraphics[width=0.315\linewidth]{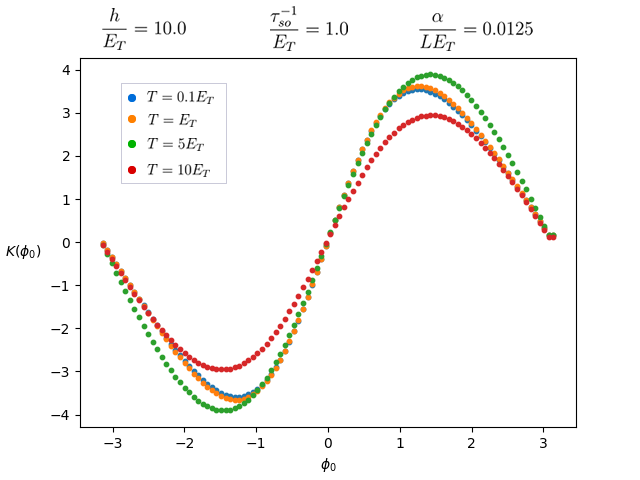}
\caption{Josephson current-phase relation of a long SNS junction computed as a function of temperature and in-plane magnetic field for the Rashba-Zeeman model.}
\label{Fig:K}
\end{figure*}

\subsection{Numerical results}

Outside the regime of the linear in magnetic fields limit, the Usadel equations must be solved numerically. For this purpose we use the methods described in Ref.~\cite{Virtanen:PRB25}. The implementation, available in Ref. \cite{Virtanen:2024}, requires discretization of the continuum action, whose saddle point is precisely the Usadel equation in the diffusive limit.

For comparison to the zero-field results, we plot in Fig.~\ref{Fig:DOS-h} density of states computed at a relatively high fields in the units of Thouless energy $h=5E_T$. To stay within the applicability domain of diffusive limit other parameters involving spin-orbit coupling and spin-relaxation were chosen accordingly as indicated on the plot. 
We observe closing of the gap in this limit. 
The main feature in the density of states is the presence of a ``well" with a width that is sensitive to $\phi_0$.
We observe weak nonreciprocity, which is displayed for two representative values of the phase $\phi_0$. On Fig.~\ref{Fig:DOS-h} we also show the sensitivity of the energy levels computed from Eq.~\eqref{eq:level_sensitivity} for the same choice of parameters in the density of states. 
Similar to the zero-field case where the level sensitivity is peaked near the gap edge, the sensitivity in this limit is peaked near the edges of the well.
It should be noted that in nonmagnetic systems the density of states is invariant under the change of sign of both the superconducting phase $\phi_0$ and magnetic field $\mathbf{H}$, namely $\nu(\epsilon,\phi_0,\mathbf{H})=\nu(\epsilon,-\phi_0,-\mathbf{H})$, which implies $V_\nu(\epsilon,\phi_0,\mathbf{H})=-V_\nu(\epsilon,-\phi_0,-\mathbf{H})$. However, changing the sign of either while keeping the other fixed does not in general imply a symmetric function. In our numerical computation we kept the direction of magnetic field fixed. 

For the generated set of density of states we also numerically computed admittance function $Q(\phi_0)$ for different temperatures and fields. The results are shown in Fig.~\ref{Fig:Q}. At very low temperatures, the dissipation $\R\Upsilon(\omega,\phi_0)\propto Q$ is noticeable only in the vicinity of the minigap closing corresponding to phases $\phi_0\sim\pm\pi$. This result is indeed expected. 
In the adiabatic regime, $\omega\tau_{in}\ll1$, and at with an increasing temperature the region of values of $\phi_0$ where $Q$ is finite broadens significantly.  The frequency dependence is governed by the Lorentizan factor in Eq. \eqref{admittance}.  At finite field we observed the appearance of the additional structure, a shoulder-like feature, which is correlated with the depletion of the density of states at intermediate energies visible at the Fig. \ref{Fig:DOS-h} in the range of energies where $\epsilon\sim h\sim E_T$. 

From Eq. \eqref{eq:J-CPR} we have also calculated the Josephson current-phase relationship. The results of this computation are displayed on Fig. \ref{Fig:K}. It is worthwhile to recall that at $T\to0$ the critical current of a long diffusive SNS junction scales with the Thouless energy and exhibits an algebraic, rather than exponential, decay with junction length, $eI_cR_N\approx 10E_T$ \cite{Dubos:PRB01,Levchenko:PRB06}, where $I_c$ is the critical current amplitude and $R_N$ is the normal state resistance of a junction. However, at any finite temperature it decays exponentially, $\propto \exp(-L/L_T)$, provided that the junction length exceeds the diffusive thermal length $L_T = \sqrt{D/T}$. For this reason, the contribution to the admittance governed by $J_s$ in Eq.~\eqref{eq:Ys} is strongly suppressed in this limit.

\begin{figure}[t!]
\includegraphics[width=\linewidth]{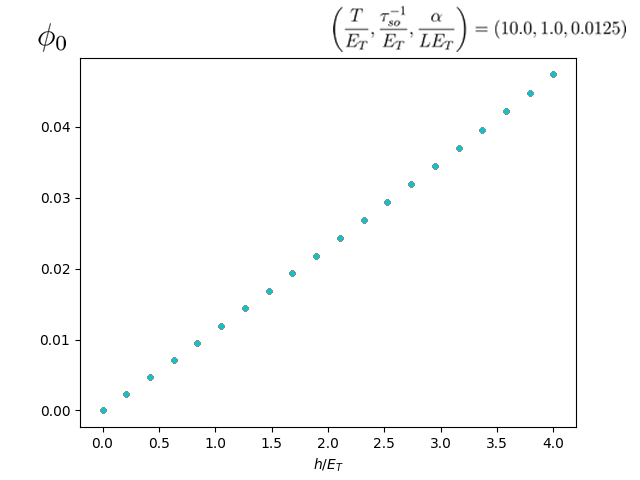}
\caption{Anomalous phase shift in the Josephson current-phase relation of a diffusive SNS junction plotted as a function of the in-plane field.}
\label{Fig:Phase}
\end{figure}

At the finite in-plane field applied to the junction the critical current is suppressed. Increasing temperature also leads to a diminishing critical current amplitude. Both of these expected features are reflected in the plots of Fig. \eqref{Fig:K}. Even though we expect to have the anomalous phase shift in the CPR and overall nonreciprocity of the Josephson current amplitude in reverse directions, these features do not clearly appear in the plot. This has to do with the magnitude of these effects in disordered systems. Applicability of the diffusive approximation of Usadel equations, $\{h,E_T\}\ll\tau^{-1}_{so}$, limits us in the allowed choice of numerical parameters. For instance, for the anomalous phase shift we can estimate $\delta\phi\sim (\alpha_R/LE_T)(\tau_{so}E_T)(h/E_T)\lesssim 5\times10^{-2}$, which is practically invisible on the plots even for the largest reasonable values of parameters in the model. To corroborate this point, we performed a more detailed numerical analysis of the anomalous phase shift. The results are displayed in Fig.~\ref{Fig:Phase}, which shows a linear scaling of $\phi_0$ with the magnetic field, with a magnitude consistent with the above estimate.

These conclusions are qualitatively consistent with the earlier similar results on the superconducting diode effect in nonreciprocal disordered systems \cite{Bergeret:EPL15,Ilic:2022,Ilic:2024,Osin:PRB24,Hasan:2025}. We can also estimate that the nonreciprocity of the critical current amplitude is even further suppressed by an additional parameter $(\Delta\tau_{el})^{3/2}\ll1$. It should be noted though that the conclusion, that the terms linear in the field can be gauged away, thus leading to a reciprocal response, arises from the one-dimensional character of the idealized model of the SNS junction. A more general situation can be envisioned, where model parameters such as the diffusion coefficients and spin-orbit coupling strengths depend on the coordinate $\mathbf{r} = (x,y)$ in the plane of the junction. Moreover, the junction itself may possess additional inhomogeneities. In these cases, the density of states cannot be expressed in the form of Eq. \eqref{nuestimateh}, which results in nonreciprocity of the response functions including the critical current.

\section{Summary}

Using a kinetic description based on the adiabatic motion of Andreev bound-state energies, we derived a general formula for the low-frequency admittance of SNS junctions [Eq.~\eqref{admittance}], expressed entirely in terms of the phase-dependent density of states $\nu(\epsilon, \phi)$. This formula reproduces known results for long diffusive junctions and remains valid for any type of SNS junction, including nonreciprocal ones, provided the frequency is sufficiently low. In nonreciprocal junctions, where both time-reversal and inversion symmetries are broken, the density of states can acquire a component that is odd in the phase difference, $\nu(\epsilon, \phi) \neq \nu(\epsilon, -\phi) $, leading to an admittance that is asymmetric with respect to the DC phase bias.

To illustrate this effect, we analyzed a diffusive planar SNS junction with an in-plane magnetic field and Rashba spin–orbit coupling in the normal region. By solving the Usadel equations, both perturbatively and through exact numerical integration, we obtained the phase-dependent density of states and the corresponding admittance. Although the frequency dependence of the admittance is restricted to the regime below the Thouless energy, the characteristic scale of the proximity-induced gap in the normal layer, it nevertheless exhibits a nontrivial Lorentzian profile with a width set by the quasiparticle inelastic relaxation rate. We find from the numerical analysis that both the admittance asymmetry and the nonreciprocity of the critical current are strongly suppressed in the diffusive limit.

The applicability of our results is not limited to SNS junctions with disordered Rashba metals. For example, diffusive junctions where the normal region is formed by the surface of a topological insulator can also be analyzed using the Usadel equation as shown in detail e.g. Ref \cite{Liu:PRB24}. Interestingly, these equations are structurally identical to Eq. \eqref{eq:Usadel-gs} used in our analysis. The only modification required is to replace the spin-relaxation time with the elastic scattering time, $\tau_{so} \to \tau_{el}$, and the Rashba velocity with the linear dispersion velocity of the surface states, $\alpha_R \to v$. Technically, these changes arise from the physics of strong spin-momentum locking of surface-state electron transport, which restricts spin relaxation to the rate of elastic scattering. Consequently, we expect that 
the S-TI-S junctions exhibit the same nonreciprocal features in their frequency-dependent admittance as the SNS junctions with Rashba spin–orbit coupling and a Zeeman field.
A promising direction for further exploration is the extension to multiterminal Josephson junctions. These systems provide an additional ingredient -- the synthetic topologies of Andreev bound states (see, e.g., Refs. \cite{Riwar:2016,Eriksson:PRB17,Xie:PRB17,Xie:PRB18}) -- which can manifest in the AC response \cite{Arnault:2021,Klees:PRL20,Heikkila:PRL24}.

\section*{Acknowledgements}

We thank Anton Andreev, Maxim Dzero, Daniel Shaffer, and Boris Spivak for useful discussions. 
This work was financially supported by the National Science Foundation (NSF), Quantum Leap Challenge Institute for Hybrid Quantum Architectures and Networks Grant No. OMA-2016136 (T. L.), 
NSF Grant No. DMR-2452658 (A. L.) and the H. I. Romnes Faculty Fellowship provided by the University of Wisconsin-Madison Office of the Vice Chancellor 
for Research and Graduate Education with funding from the Wisconsin Alumni Research Foundation.

\section*{Data availability} 

All data presented in the figures were generated from the publicly available package that can be downloaded from \cite{Virtanen:2024}.

\bibliography{nonRecAC}{}

\end{document}